# Quantum Correlations and Superluminal Interaction

Alexander V. Belinsky[*], and Michael H. Shulman[§]

[*]*Moscow State University, Department of Physics* 119992 Russia
*E-mail: belinsky@physics.msu.ru, belinsky@inbox.ru*

[§]*Independent Researcher  Moscow, Russia*
*E-mail: shulman@dol.ru*

PACS numbers: 03.65.Ud

The paper attempts to solve the well known conflict between Relativity and Quantum Mechanics. Accordingly to our concept the instant correlations between the EPR-partners can be explained by some oscillations existence whose propagation velocity may overcome the velocity of light in vacuum. In such process any irreversible transport of energy and information is absent, so the relativistic causality principle is satisfied.

## I. INTRODUCTION

There exists now the common opinion that Relativity interdicts any physical interaction propagating with a superluminal speed over any (arbitrary large) distances (Einstein – Podolsky – Rosen Paradox). This was confirmed by a number of experiments (see, e.g. [1, 2]), hence the insuperable conflict arises between QM and Relativity. In 1990 John Bell said [3]:

> We have the statistical predictions of quantum mechanics, and they seem to be right. The correlations seem to cry out for an explanation, and we don't have one.

Further, he expressed a hope:

> …here, I think we have a temporary confusion. It's true that it is sixty years old, but on the scale of what I hope will be human existence, that's a very small time. I think the problems and puzzles we are dealing with here will be cleared up, and we will look back on them with the same kind of superiority, our descendants will look back on us with the same kind of superiority as we now are tempted to feel when we look at people in the late nineteenth century who worried about the ether. And Michelson-Morley .., the puzzles seemed insoluble to them. And came Einstein in nineteen five, and now every schoolboy learns it and feels .. superior to those old guys. Now, it's my feeling that all this action at a distance and no action at a distance business will go the same way. But someone will come up with the answer, with a reasonable way of looking at these things. If we are lucky it will be to some big new development like the theory of relativity. Maybe someone will just point out that we were being rather silly, and it won't lead to a big new development.

There exist many publications in which different approaches and interpretations were discussed. It is evident that some *fundamental* statement should be introduced either into Relativity, or into QM, or into both. We propose below one of the possible solutions that specifies the relativistic causality conception in Relativity.

## II. STANDARD CAUSALITY PRINCIPLE IN RELATIVITY

Without loss of generality one can formulate the modern relativistic causality concept as the statement that a *signal[1] and energy* superluminal propagation is impossible. Further, this statement can be examined with a freely moving particle having a mass at rest m and flying from a point 1 to a point 2.

Let us now choose the comoving relativistic frame where dx=0, and only temporal coordinate changes. So, the 4-distance between these events is always positive:

$$ds^2 = c^2dt^2 - dx^2 = c^2dt^2 > 0. \qquad (1)$$

One can associate this pair of 4-events (which are ordered in spacetime) of above particle with the relativistic invariant expression for a free particle action [4]:

$$S_{12} = -mc \int_1^2 ds = -mc^2 \int_1^2 d\tau = -mc^2 T_{12}, \qquad (2)$$

where ds is an element of 4-distance, $d\tau$ is a differential of the *proper* time (i.e., of the time in the comoving reference frame), $T_{12}$ is the proper time between events 1 and 2. It is clear that for a chain of successive particle displacements from 1 to 2, 2 to 3, ..., N-1 to N the total action will be equal to the partial contributions sum: $S_{1N} = S_{12} + S_{23} + \ldots + S_{N-1, N}$. It is also clear that a sign change corresponds to the time *reversal*, i.e., to the energy and information transfer from 4-point 2 to the 4-point 1.

One usually assumes that the action $S_{12}$ is here a real valued quantity and the proper time $T_{12}$ is timelike, because energy cannot propagate with a superluminal speed.

This consideration also holds for a photon which hasn't a mass at rest. In fact, a photon registration (as an information carrier) is associated with absorption of its momentum, so we can again talk about some energy transport from one point to another.

## III. TO CUT GORDIAN KNOT

Let us cut the Gordian knot of the conflict between Relativity and QM using a convenient specification of the relativistic causality principle. Such specification does not violate the basic aspects of this principle. Contrary, it expands them and allows to better understand several fundamental representations of QM.

As we noted in the beginning of the paper QM leads to a possibility of a quantum state instant reduction (collapse). It is clear that "non-locality" is equivalent to a possibility of arbitrary large velocities of some interaction (may be, *interference*) whose essence requires some explicit and clear physical description.

Let us assume that a superluminal motion is *possible* and consider what follows from this assumption. In this case the 4-distance T in Eq. (2) will be (as it is well-known, see [4]), an *imaginary* one, so the action S will be *imaginary* too.

The imaginary durations, coordinates and other physical quantities are used in physics for a long time, and especially in quantum physics. In the Heisenberg's seminal work (1925) its author started by representation of a quantum particle coordinate and momentum as infinite sets of complex valued terms (see [5]): he associated transition amplitudes and energies between two atomic states with amplitudes and frequencies of individual complex terms. After that it turned out that several additional limitations had to be introduced in order to have *real* values of physical quantities. As a result the new kind of multiplication turned out to be a non-commutative one. This fact as well as representation of physical quantities by sets of time-dependent complex numbers was very striking; Born was puzzled. Several days ago he recalled an algebraic theory which he had learnt in Breslau. Born recognized that Heisenberg's sets were precisely those matrices with which he had become acquainted when in 1903, as a young student at the University of Breslau, he attended lectures on algebra and analytical geometry. Thus, in physical community the correctness of complex values use in quantum physics (as well as in Relativity) is not now debated.

---

[1] The signal registration is always associated with the energy absorption by a detector.

However, what such the situation physically means? We propose a surprising answer to this question. For example, complex quantities are extensively used to describe active (aperiodic) and reactive (purely periodic) harmonic electrical processes; the imaginary values correspond only to the oscillating quantities. In this theory the complex values generally represent generally electrical tensions (analogues of coordinates), electrical currents (analogues of momentums), and powers. E.g., *apparent* (complex) power in a circuit (per cycle) one can represent as [6]:

$$D = \frac{U_m I_m}{2} \exp[\,j\varphi\,] = \frac{U_m I_m}{2} (\cos\varphi + j\sin\varphi), \qquad (3)$$

where $U_m$ - the harmonic tension magnitude, $I_m$ - the harmonic current magnitude, $\varphi$ - the phase shift between tension and current harmonics, $j$ - the imaginary unit. Here $\cos\varphi$ corresponds to the active (dissipated in the circuit) power, and $\sin\varphi$ corresponds to the reactive (oscillating) power.

Let us introduce the new variable

$$\psi = \sqrt{\frac{U_m I_m}{2}} \exp[\,j(-\varphi/2)], \quad \psi^* = \sqrt{\frac{U_m I_m}{2}} \exp[\,j(\varphi/2)], \qquad (4)$$

then one can easily see that the power modulo in the circuit is

$$|D| = \psi^*\psi. \qquad (5)$$

Further, let us remember the expression for the probability density computed with the wave function $\psi$ (of course, this quantity differs from above electrical one):

$$p = \psi^*\psi. \qquad (6)$$

These two relationships are similar, aren't they?

Therefore, we want claim that Heisenberg implicitly separated in QM two types of processes. The processes of the first type correspond to classical analogues – we can directly measure them (let us call them conditionally as relatively "slow" ones), they are described as aperiodic. The processes of the second type are described as imaginary ones, they correspond to the purely oscillating terms and very high frequencies. These frequencies correspond to a particle *energy at rest* (one uses the special word "zitterbewegung" for Dirac's electron [7]). Note, the *mean energy* transport in such the process is zero.

Thus, we can now say that:

- the purely oscillating (in time) processes play the principally important role in physical theories;
- the complex numbers are convenient and natural tool to describe the oscillation processes;

- while a physical process is purely oscillating (i.e., does not contain any aperiodic component) the energy *transport* from one spatial point to another is absent. However, this generally does not means that *any* physical interference is absent too. In this case irreversible energy and information transport is really absent, the *mean* value of the transporting (to any side) energy is *zero* that completely corresponds to the relativistic causality concept.

This is the required generalization of the relativistic causality concept that allows us to obtain the coincidence between Relativity and QM.

## IV. QUANTUM CORRELATIONS

What does the expression "quantum correlations between states" mean? In 1935 E. Shrödinger invented the scintific word "entanglement", and Einstein-Podolsky-Rosen wrote the paper [8], from wich it became clear: according to QM any entangled EPR-particle (photon, electron, …) arbitrary distant from another one remains sensitive to any state perturbation of this second particle, and vice-versa. If one performs a number of experiments on single pairs he reveals a statistical correspondence measure between the pair particles features (in several extreme cases such the measure may become deterministic [9] dependently on a measurement configuration). However, the connection between entangled particles cannot be relativistically causal as we noted in the beginning of the paper.

Now we can rely on our new idea: a more weak (not causal one!) physical connection type is possible that is not associated with any *irreversible* energy transport from one 4-point to another. Contrary, if our idea holds, then such the connection type is specified by a purely oscillating process and accordingly a superluminal propagation velocity.

Let us consider an entangled quantum state of EPR-pair. A standard description deals with Alisa and Bob who measure their own particles. Alisa can reveal its particle in the state $a$ or $A$, while Bob can reveal its particle in the state $b$ or $B$. Since the particles are entangled only two combinations ($ab$ and $AB$) are physically possible from four ones ($ab$, $aB$, $Ab$ и $AB$). I.e., common *entangled* state may be represented as

$$\left|\psi\right\rangle = \frac{1}{\sqrt{2}}\left(\left|ab\right\rangle + \left|AB\right\rangle\right). \tag{7}$$

In this case the density matrix $\rho$ is:

$$\rho = \begin{pmatrix} 1/2 & 0 & 0 & 1/2 \\ 0 & 0 & 0 & 0 \\ 0 & 0 & 0 & 0 \\ 1/2 & 0 & 0 & 1/2 \end{pmatrix}. \tag{8}$$

Thus, the system can be found in the state $\left|ab\right\rangle$ or $\left|AB\right\rangle$ with the same probability 1/2 (terms indexed as [1,1] and [4,4]), the correlation between these states are maximal (terms indexed as [1,4] and [4,1]). This result holds *before* the measurement, while *after* measurement we obtain the matrix $\mu$ that has only main diagonal terms:

$$\mu = \begin{pmatrix} 1/2 & 0 & 0 & 0 \\ 0 & 0 & 0 & 0 \\ 0 & 0 & 0 & 0 \\ 0 & 0 & 0 & 1/2 \end{pmatrix}. \tag{9}$$

So, the measurement (and the collapse due to it) leads to the transition $\rho \rightarrow \mu$, i.e., the non-diagonal terms vanish.

It seems that in spite of a distance between the particles some oscillating energy exchange occurs[2], and the pair itself exists as a superposition of two basic states. When we perform the measurement on one of them (or at its re-entanglement with some another system) this superposition becomes to be destroyed, the terms indexed as [1,4] and [4,1] vanish, the decoherence occurs [10] (see also similar description of the Shrödinger' cat evolution in [11]).

In fact, as it is noted in [10] the measurement process represents the interaction between measured particle and a measuring apparatus or some environment, and because of that the particle and apparatus (environment) states become entangled, a quantum correlations appears between them. In the same time the entanglement and correlations between the two original particles destroy. If such a picture holds, one may formulate and investigate the problem of the *superposition reduction velocity*. Indeed, in [11] several analytic and numeric estimates are given for the decoherence time, typically near $10^{-23}$ s.

Let us add a little about some possible information exchange without energy exchange. When we consider a classical periodical process (e.g., alternating electrical current) even if an active loss is absent one can talk about the zero energy transport only on average. However, if we consider a duration that is less than the cycle time, then the qualitative measure of the transported energy can always be determined not only theoretically, but practically.

This situation becomes in principle different when we consider the quantum domain and, particularly, so called vacuum zero-point fluctuations. Such fluctuations correspond to the ground electromagnetic field state; the field strength *averaged over space* and the photon number are equal to zero in it, however, an averaged *square* of the field strength differs from zero. It becomes apparent implicitly in the Lamb shift [12], Casimir effect [13], seeded field of optical parametric oscillator (see, for example, [14, 15]).

We are sure that this vacuum zero-point fluctuations energy cannot be extracted directly (for example, at a measurement) because it corresponds to the *lowest* possible energy level. So, in this case any irreversible and non-zero energy exchange cannot be realized in principle.

## V. SUPERLUMINAL VELOCITIES

In Relativity one believes impossible to overcome the velocity of light, as it is stated by the velocities addition formula (see, for example, [4]). Let reference frame $K'$ move relative to reference frame $K$ with the speed $V$. Further, let $v$ be the velocity in the reference frame $K$ and $v'$ be the same particle velocity in the reference frame $K'$ (for the sake of simplicity we assume that all velocities have a common direction). Then

$$v = \frac{v' + V}{1 + \frac{v'V}{c^2}}, \tag{10}$$

---

[2] The photon oscillation frequency is probably equal to its own frequency however, a usual connection with the velocity of light is impossible (the nature of such the oscillation is unknown). The electron oscillation frequency (as we believe) corresponds to its de-Broglie frequency [7].

where $c$ is velocity of light. It is easy to see that the sum of two velocities that are less than $c$ really cannot be more than $c$.

Let us now remember one instructive story. In 1916 K. Schwarzschild found the Einstein's field equation solution describing a black hole. The Schwarzschild metric using Schwarzschild coordinates is given by (see at [16], [17]):

$$ds^2 = (1 - \frac{r_s}{r})c^2 dt^2 - \frac{dr^2}{(1 - \frac{r_s}{r})} - r^2 (\sin^2\theta d\varphi^2 + d\theta^2), \qquad (11)$$

where $s$ is 4-distance, $c$ is velocity of light, $t$, $r$, $\theta$, $\varphi$ are the Schwarzschild coordinates, $r_s$ =2GM/c$^2$ is Schwarzschild radius ($M$ is the black hole mass). It is easy to see that at $r = r_s$ the singularity appears in the second term of (11). This physically means that for a remote observer the speed of an object falling into BH tends to 0 as $r$ approaches the event horizon. The object appears to have slowed as it gets nearer the event horizon and halted at the event horizon. While one intersects the event horizon transition the terms $dt$ and $dr$ change their signs and these time and radial coordinates seem to be exchanged for their roles.

However, 17 years later G. Lemaître [16] was the first to show that this is not a real physical singularity but simply a manifestation of the fact that the static Schwarzschild coordinates cannot be realized with material bodies inside the gravitational radius. Such singularity can be eliminated using the coordinate system transformation:

$$d\tau = dt + \sqrt{\frac{r_s}{r}} \frac{1}{1 - \frac{r_s}{r}} dr, \quad d\rho = dt + \sqrt{\frac{r}{r_s}} \frac{1}{1 - \frac{r_s}{r}} dr, \qquad (12)$$

where the metric is:

$$ds^2 = d\tau^2 - \frac{r_s}{r} d\rho^2 - r^2 (\sin^2\theta d\varphi^2 + d\theta^2), \qquad r = [\frac{3}{2}(\rho - \tau)]^{2/3} r_s^{1/3}. \qquad (13)$$

As we can see, in this coordinate system the new time and radial coordinates are both expressed through old corresponding coordinates, so such the transformation is not trivial, but now the singularity is exactly in the BH geometrical center as should be expected. It is possible now to describe a process inside of black hole. The striking situation appears as follows: in one reference frame the event horizon (and light barrier) cannot be surmounted while in second one this barrier can be surmounted (in one direction only) by a falling object.

Indeed, in contrast to the Schwarzschild coordinate system, in the new system the falling object velocity becomes equal to $c$ at the event horizon and then it can be expressed as $(dr/dt_{pr})_{obj} = -c\sqrt{r_s/r}$ (here $t_{pr}$ is proper time of a comoving observer), i.e., it increases up to infinity at the center (at the singularity). In this case we should express the light velocity as it was *measured in the same reference frame*; so it is not now equal to the constant value $c$, but is $(dr/dt_{pr})_{light} = -c(1 + \sqrt{r_s/r})$, hence, it also increases unlimitedly; however, it remains always more (by modulo) than the falling body velocity [17].

Next investigations are associated with W. Rindler who searched for coordinate system where a comoving observer experiences the uniform acceleration (see [18]). It is clear that he will obtain the velocity of light (exactly like falling into black hole), and ...? That's right, a *remote* observer will not be able to trace the next way of the comoving observer. However, Rindler proposed the new coordinate system that consists in two "wedges" and do not cover the entire Minkowski space–time; the wedges are limited by the events horizons. The author of [18] writes about the horizons and emergent Hawking emission:

In fact, horizons can be made in a simple kitchen experiment. Just let water flow out of the tap onto a metal surface <…> Inside a certain ring the water surface is very smooth, but outside waves are appearing. Where the stream from the tap hits the metal, the water flows faster than the wave velocity. Then the water flows outwards and gets slower. Waves cannot enter the region where the water exceeds their velocity, but they are formed at the critical radius where the water has reached the wave velocity. In fluid mechanics, this phenomenon is known as the *hydraulic jump*. Seen from an astrophysics perspective, the hydraulic jump resembles a *white hole*, an object that nothing can enter. White holes are time–reversed black holes — if we run a movie of the kitchen experiment backwards in time and turn it upside down, the water seems to flow towards a water fall.

<…> Imagine the two observers are on two conjugate Rindler trajectories, one on a space–time hyperbola with fixed positive proper acceleration, (on the right side of the Rindler wedge) and the second observer on a trajectory with a negative proper acceleration, (on the left side of the Rindler wedge) <…> Both observers are part of the same space–time coordinate system, the Rindler chart <…> Instead of the Minkowski vacuum, the two accelerated observers witness the Einstein–Podolski–Rosen state.

Both accelerated observers individually perceive the Minkowski vacuum as thermal radiation, but if they compared their records after the acceleration stage they would notice that the thermal photons were always correlated. The recoil kicks that slow them down have been synchronized! So, apparently random events caused by the quantum vacuum have been correlated across space; they have been entangled…

The two partners in observation are on the two sides of the Rindler wedge, they are separated by a *horizon*. The horizon is the space–time surface where the acceleration is so strong that an observer would instantly reach the speed of light.

Thus, we can suppose that light barriers and superluminal velocities may be considered in physics; however, in order to do it several non-trivial coordinate systems have to be used.

## VI. CONCLUSION

We suggest that the representation of an imaginary (reactive) energy allows us to better understand not only non-local correlations between particles of the EPR pair but other phenomena too, e.g., particle tunneling through a potential barrier, see, particularly, the work [19], where the authors point out that the peak of a tunneling wave packet may indeed appear on the far side of a barrier sooner than if it had been travelling at the vacuum speed of light. No signal can be sent with these smooth wavepackets, however; only a small portion of the leading edge of the incident Gaussian is actually transmitted. The relativistic causality is thus not violated by these nonlocal effects. See also [20] where its author writes: a tunneling particle spends *purely imaginary* time on a barrier region.

J. Bell [3] talked about A. Einsteins's argument on the QM incompleteness:

In 1935 he invented an extremely powerful argument, for this position, based on another hypothesis which most people who have not met these phenomena before would accept; the hypothesis of no action at a distance, which is sometimes called local causality or just locality. And he said that there are situations where this hypothesis implies determinism. So in this argument determinism was no longer a hypothesis, but a theorem, but with locality as the axiom.

Bell himself discovered a violation of such the locality axiom just during the OM completeness investigation. Thus, if one recognizes the numerous experiments results, he has to accept a possibility of a quantum state vector *instant* propagation (however, at limited velocity of an *active* energy and information), and start to investigate a concrete theory that could adequately describe such a process.

There exist several works in which just such purposes are declared. For example, in the paper [21] a diffusion wave model based on the parabolic type equation for thermal vacuum field is proposed. Mathematically, diffusion waves are characterized by the peculiarity that the time derivative in their defining equation is only of first order. They are wave-like disturbances characterized by coherent, and always driven, oscillations along diffusing energy or particles. They show an infinite speed of propagation of disturbances along entire domains [21]. Although the proposed model can be debated, the paper is finished by a typical statement that one of the most exciting questions to be tackled in the future will be how all of this can be described in relativistic terms.

Let us summarize the new proposed representations.

1. We believe that the Relativity interdiction to overcome the velocity of light in vacuum is not *absolute*: it is true for a remote observer, however, a comoving one is able to overcome event horizon (in a single direction), quantum correlated photons may appear on both sides.

2. Due to this we can physically consider several purely alternating processes propagating with a superluminal velocity in a remote frame. In such processes any transport of average energy is absent, because of that the relativistic causality is not violated. However, some physical interference between two EPR-particles may be possible, because averaged *square* energy is positive (non-zero); as we believe, it leads to a more weak connection between them than causal interaction. It is interesting that mathematically the alternating processes are adequately described by the complex numbers that correspond with imaginary 4-distances and 4-action in Minkowski space.

3. We share the statement that an entanglement between two EPR-particles can be destroyed at a *measurement* on one from them because of a corresponding re-entanglement with an apparatus or environment [10]. We suggest that new investigations progress has to be connected to the searching for state collapse rate in EPR-experiments.